\documentclass[12pt]{article}

\usepackage{amsmath,amssymb,amscd,amsbsy,amsgen,amsopn,amstext,
amsxtra}

\usepackage[dvips]{graphicx}

\begin{document}

\vskip 0.5 truecm

\begin{center}
{\Large{\bf Berry's phases and topological properties in the 
Born-Oppenheimer approximation\footnote{Talk given at the International Symposium on the Foundations of Quantum Mechanics, Hatoyama, Japan, August 22-25, 2005 (to be published in Proceedings).}}}
\end{center}
\vskip .5 truecm
\centerline{\bf  Kazuo Fujikawa }
\vskip .4 truecm
\centerline {\it Institute of Quantum Science, College of 
Science and Technology}
\centerline {\it Nihon University, Chiyoda-ku, Tokyo 101-8308, 
Japan}
\vskip 0.5 truecm

\begin{abstract}
The level crossing problem is neatly formulated by the second quantized formulation, which exhibits a hidden local gauge symmetry. The analysis of geometric phases is reduced to a simple diagonalization of the Hamiltonian. If one diagonalizes the geometric terms in the infinitesimal neighborhood of level crossing, the geometric phases become trivial (and thus no monopole singularity) for arbitrarily large but finite time interval $T$. The topological proof of the Longuet-Higgins' phase-change rule, for example, thus fails in the practical Born-Oppenheimer approximation where $T$ is identified with the period of the slower system. The crucial difference between the Aharonov-Bohm phase and the geometric phase is explained. It is also noted that the gauge symmetries involved in the  adiabatic and non-adiabatic geometric phases are quite different.
\end{abstract}


\section{Introduction}
The geometric phases revealed the importance of 
hitherto un-recognized phase factors in the adiabatic 
approximation\cite{berry,simon,stone,higgins,kuratsuji,berry2}. 
It may then be 
interesting to investigate how 
those phases behave in the exact formulation. We formulate the 
level crossing problem by using the second quantization 
technique, which works both in the path integral and operator 
formulations\cite{fujikawa,fujikawa2,fujikawa3}. In this 
formulation, the analysis of geometric 
phases is reduced to the familiar diagonalization of the  
Hamiltonian. Also, a hidden local gauge
 symmetry replaces the  notions of parallel transport and 
holonomy.

When one diagonalizes the 
Hamiltonian in a very specific limit, one recovers the 
conventional geometric phases defined in the adiabatic 
approximation. If one diagonalizes the Hamiltonian in the other 
extreme limit, namely, in the infinitesimal neighborhood of 
level crossing for any fixed finite time interval $T$, one can 
show that the geometric phases become trivial and thus no 
monopole-like singularity. At the level crossing point, the 
conventional energy eigenvalues become degenerate but the 
degeneracy is lifted if one diagonalizes the geometric terms. 
Our analysis shows that the topological 
interpretation\cite{stone,berry} of geometric phases such as 
the topological proof of the Longuet-Higgins' phase-change 
rule\cite{higgins} fails in the practical Born-Oppenheimer 
approximation where $T$ is identified with the period of the 
slower system. This analysis shows that the topological properties of the geometric phase and the Aharonov-Bohm phase are quite different. 

Also, the difference between gauge symmetries for adiabatic phase and "non-adiabatic phase" by  Aharonov-Anandan\cite{aharonov} becomes quite clear in this formulation.
 
\section{Second quantized formulation}

We start with the generic hermitian Hamiltonian 
$\hat{H}=\hat{H}(\hat{\vec{p}},\hat{\vec{x}},X(t))$
for a single particle theory in a slowly varying background 
variable $X(t)=(X_{1}(t),X_{2}(t),...)$. 
The path integral for this theory for the time interval
$0\leq t\leq T$, which is taken to be the period of the slower 
background system, in the second quantized 
formulation is given by 
\begin{eqnarray}
&&\int{\cal D}\psi^{\star}{\cal D}\psi
\exp\{\frac{i}{\hbar}\int_{0}^{T}dtd^{3}x[{\cal L}
] \}
\nonumber
\end{eqnarray}
where
\begin{eqnarray}
{\cal L}&=&\psi^{\star}(t,\vec{x})i\hbar\frac{\partial}{\partial t}
\psi(t,\vec{x})-\psi^{\star}(t,\vec{x})
\hat{H}(\frac{\hbar}{i}\frac{\partial}{\partial\vec{x}},
\vec{x},X(t))\psi(t,\vec{x}).
\end{eqnarray}
We then define a complete set of eigenfunctions
\begin{eqnarray}
&&\hat{H}(\frac{\hbar}{i}\frac{\partial}{\partial\vec{x}},
\vec{x},X(t))v_{n}(\vec{x},X(t))={\cal E}_{n}(X(t))v_{n}(\vec{x},X(t)), \nonumber\\
&&\int d^{3}x v^{\star}_{n}(\vec{x},X(t))v_{m}(\vec{x},X(t))
=\delta_{n,m}
\end{eqnarray}
and expand 
\begin{eqnarray}
\psi(t,\vec{x})=\sum_{n}b_{n}(t)v_{n}(\vec{x},X(0)).
\end{eqnarray}
The path integral for the time interval $0\leq t\leq T$ in the 
second quantized formulation is given by  
\begin{eqnarray}
Z&=&\int \prod_{n}{\cal D}b_{n}^{\star}{\cal D}b_{n}\exp\{\frac{i}{\hbar}\int_{0}^{T}dt[
\sum_{n}b_{n}^{\star}(t)i\hbar\frac{\partial}{\partial t}
b_{n}(t)\nonumber\\
&&+\sum_{n,m}b_{n}^{\star}(t)
\langle n|i\hbar\frac{\partial}{\partial t}|m\rangle
b_{m}(t)-\sum_{n}b_{n}^{\star}(t){\cal E}_{n}(X(t))b_{n}(t)] \}
\end{eqnarray}
where the second term in the action stands for the term
commonly referred to as Berry's phase\cite{berry} and its 
off-diagonal generalization. 
The second term  is defined by
\begin{eqnarray} 
&&\langle n|i\hbar\frac{\partial}{\partial t}|m\rangle\equiv
\int d^{3}x v^{\star}_{n}(\vec{x},X(t))
i\hbar\frac{\partial}{\partial t}v_{m}(\vec{x},X(t)).
\end{eqnarray}

In the operator formulation of the second quantized theory,
we thus obtain the effective Hamiltonian
\begin{eqnarray}
\hat{H}_{eff}(t)&=&\sum_{n}\hat{b}_{n}^{\dagger}(t)
{\cal E}_{n}(X(t))\hat{b}_{n}(t)-\sum_{n,m}\hat{b}_{n}^{\dagger}(t)
\langle n|i\hbar\frac{\partial}{\partial t}|m\rangle
\hat{b}_{m}(t).
\end{eqnarray} 

When we define the Schr\"{o}dinger picture 
$\hat{{\cal H}}_{eff}(t)$ by replacing all
$\hat{b}_{n}(t)\rightarrow \hat{b}_{n}(0)$ in $\hat{H}_{eff}(t)$ 
we can show\cite{fujikawa,fujikawa2}
\begin{eqnarray}
&&\langle n(T)|T^{\star}\exp\{-\frac{i}{\hbar}\int_{0}^{T}
\hat{H}(\hat{\vec{p}}, \hat{\vec{x}},  
X(t))dt \}|n(0)\rangle=\langle n|T^{\star}\exp\{-\frac{i}{\hbar}\int_{0}^{T}
\hat{{\cal H}}_{eff}(t)
dt\}|n\rangle.
\end{eqnarray}
Both-hand sides of this formula are exact, but the difference is 
that the geometric terms, both of diagonal and off-diagonal, 
are explicit in the second quantized formulation on the 
right-hand side. The state vectors in the second 
quantization are defined by 
$|n\rangle=\hat{b}_{n}^{\dagger}(0)|0\rangle$,
and the state vectors in the first quantized states by
(2). If one retains only the diagonal elements in 
this formula (7), one
recovers the conventional adiabatic formula\cite{kuratsuji}
\begin{eqnarray}
\exp\{-\frac{i}{\hbar}\int_{0}^{T}dt 
[{\cal E}_{n}(X(t))
-\langle n|i\hbar\frac{\partial}{\partial t}|n\rangle]\}.
\end{eqnarray}

The above formula (7) represents the essence of geometric 
phases: If 
one performs an exact evaluation one does not obtain a clear 
physical picture of what is going on. On the other hand, if one
makes an adiabatic approximation one obtains a clear universal 
picture.

 The path integral formula (4) is based 
on the expansion (3), and the starting theory depends only on 
the field variable $\psi(t,\vec{x})$, not on  $\{ b_{n}(t)\}$
and $\{v_{n}(\vec{x},X(t))\}$ separately. This fact shows that 
our formulation contains a hidden local gauge symmetry 
\begin{eqnarray}
&&v^{\prime}_{n}(\vec{x},X(t))=
e^{i\alpha_{n}(t)}v_{n}(\vec{x},X(t)),\nonumber\\
&&b^{\prime}_{n}(t)=
e^{-i\alpha_{n}(t)}b_{n}(t)
\end{eqnarray}
where the gauge parameter $\alpha_{n}(t)$ is a general 
function of $t$. By using this gauge freedom, one can choose the 
phase convention of the basis set $\{v_{n}(\vec{x},X(t))\}$ at one's will such 
that the analysis of geometric phases becomes simplest. 
From the view point of hidden local symmetry, the formula (8) is 
a result of the specific choice of eigenfunctions 
$v_{n}(\vec{x},X(0))=v_{n}(\vec{x},X(T))$ in the gauge invariant 
expression
\begin{eqnarray}
&&v_{n}(\vec{x}; X(0))^{\star}v_{n}(\vec{x};X(T))
\exp\{-\frac{i}{\hbar}\int_{0}^{T}[{\cal E}_{n}(X(t))
-\langle n|i\hbar\frac{\partial}{\partial t}|n\rangle]dt\}.
\end{eqnarray}
This hidden
local symmetry replaces the notions of paralell transport and holonomy
in the analyses of geometric phases, and it works not only for 
cyclic but also for non-cyclic evolutions\cite{fujikawa3}.

\section{Level crossing problem}

For a simplified two-level problem, the Hamiltonian is defined by the matrix in the neighborhood of level crossing~\cite{fujikawa} 
\begin{eqnarray}
h(X(t))
&=&\left(\begin{array}{cc}
            E(t)&0\\
            0&E(t)
            \end{array}\right)
      +g \sigma^{l}y_{l}(t)
\end{eqnarray}
after a suitable re-definition of the parameters by taking linear 
combinations of  $X_{k}(t)$. Here $y_{l}(t)$ stands for the background 
variable and  $\sigma^{l}$ for the Pauli matrices, and $g$ is a 
suitable (positive) coupling constant.

The eigenfunctions in the present case are given by 
\begin{eqnarray}
&&v_{+}(y)=\left(\begin{array}{c}
            \cos\frac{\theta}{2}e^{-i\varphi}\\
            \sin\frac{\theta}{2}
            \end{array}\right), \ \ \
v_{-}(y)=\left(\begin{array}{c}
            \sin\frac{\theta}{2}e^{-i\varphi}\\
            -\cos\frac{\theta}{2}
            \end{array}\right)
\end{eqnarray}
by using the polar coordinates, 
$y_{1}=r\sin\theta\cos\varphi, \ y_{2}=r\sin\theta\sin\varphi,
\ y_{3}=r\cos\theta$. Note that, by using hidden local symmetry, our eigenfunctions are chosen to be periodic under a $2\pi$ rotation around 3-axis, which is quite different from a 
$2\pi$ rotation of a spin-1/2 wave function. If one defines
\begin{eqnarray} 
v^{\dagger}_{m}(y)i\frac{\partial}{\partial t}v_{n}(y)
=A_{mn}^{k}(y)\dot{y}_{k}  \nonumber
\end{eqnarray}
where $m$ and $n$ run over $\pm$, we have
\begin{eqnarray}
A_{++}^{k}(y)\dot{y}_{k}
&=&\frac{(1+\cos\theta)}{2}\dot{\varphi}
\nonumber\\
A_{+-}^{k}(y)\dot{y}_{k}
&=&\frac{\sin\theta}{2}\dot{\varphi}+\frac{i}{2}\dot{\theta}
=(A_{-+}^{k}(y)\dot{y}_{k})^{\star}
,\nonumber\\
A_{--}^{k}(y)\dot{y}_{k}
&=&\frac{1-\cos\theta}{2}\dot{\varphi}.
\end{eqnarray}

The effective Hamiltonian (6) is then given by 
\begin{eqnarray}
\hat{H}_{eff}(t)&=&(E(t)+ g r(t))\hat{b}^{\dagger}_{+}
\hat{b}_{+}
+(E(t)- g r(t))\hat{b}^{\dagger}_{-}\hat{b}_{-}
-\hbar \sum_{m,n}\hat{b}^{\dagger}_{m}A^{k}_{mn}(y)\dot{y}_{k}
\hat{b}_{n}.
\end{eqnarray} 
with $r(t)=\sqrt{y^{2}_{1}+y^{2}_{2}+y^{2}_{3}}$. The point
$r(t)=0$ corresponds to the level crossing. In the adiabatic approximation, one neglects the off-diagonal
terms in the last geometric terms, which is justified for 
$Tg r(t)\gg \hbar\pi$,
where $\hbar\pi$ stands for the magnitude of the geometric term 
times $T$. The adiabatic formula (8) then gives the familiar
result
\begin{eqnarray}
&&\exp\{i\pi(1-\cos\theta) \}\times\exp\{-\frac{i}{\hbar}\int_{C(0\rightarrow T)}dt
[E(t)- g r(t)] \}
\end{eqnarray}
for a $2\pi$ rotation in $\varphi$ with fixed $\theta$, for 
example. 

\normalsize
\begin{figure}[!htb]
 \begin{center}
    \includegraphics[width=10.9cm]{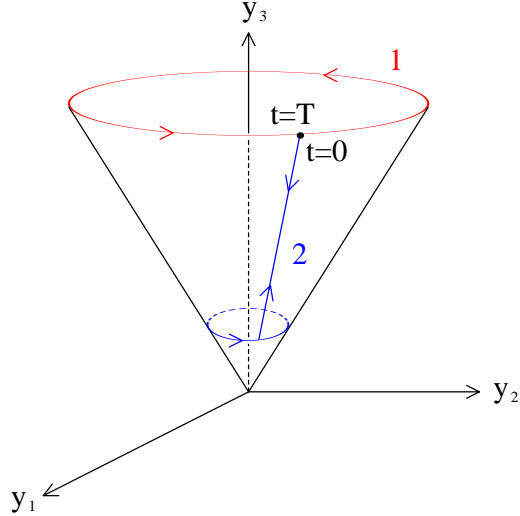} 
       \end{center}
\vspace{-9mm}      
 \caption{ 
 The path 1 gives the conventional 
geometric phase for a fixed finite $T$, 
whereas the path 2 gives a trivial geometric phase for a fixed finite $T$. Note that both of the paths cover the 
same solid angle $2\pi(1-\cos\theta)$.  }
\end{figure}

\vspace{1mm}

To analyze the behavior near the level crossing point, we 
perform a unitary 
transformation 
$\hat{b}_{m}=\sum_{n}U(\theta(t))_{mn}\hat{c}_{n}$
where $m,n$ run over $\pm$ with
\begin{eqnarray}
U(\theta(t))=\left(\begin{array}{cc}
            \cos\frac{\theta}{2}&-\sin\frac{\theta}{2}\\
            \sin\frac{\theta}{2}&\cos\frac{\theta}{2}
            \end{array}\right),
\end{eqnarray}
which diagonalizes the geometric terms and the above effective Hamiltonian (13) is written as
\begin{eqnarray}
\hat{H}_{eff}(t)&\simeq& (E(t)+gr\cos\theta)
\hat{c}^{\dagger}_{+}\hat{c}_{+}
+(E(t)-gr\cos\theta)\hat{c}^{\dagger}_{-}\hat{c}_{-}
-\hbar\dot{\varphi} \hat{c}^{\dagger}_{+}\hat{c}_{+}
\end{eqnarray}
in the infinitesimal neighborhood of the level crossing point,
namely, for sufficiently close to the origin of the parameter 
space $(y_{1}(t), y_{2}(t), y_{3}(t) )$ but 
$(y_{1}(t), y_{2}(t), y_{3}(t))\neq (0,0,0)$.
To be precise, for any given {\em fixed} time interval $T$, 
we can choose in
the infinitesimal neighborhood of level crossing
$T gr(t)\ll T\hbar\dot{\varphi}\sim 2\pi\hbar$.
In this new basis, the geometric phase appears only for
 the mode $\hat{c}_{+}$ which gives rise to a phase factor
\begin{eqnarray}
\exp\{i\int_{C} \dot{\varphi}dt \}=\exp\{2i\pi \}=1,
\end{eqnarray}
and thus no physical effects. In the infinitesimal neighborhood 
of level crossing, the states spanned by 
$(\hat{b}_{+},\hat{b}_{-})$ are transformed to a linear 
combination of the 
states spanned by $(\hat{c}_{+},\hat{c}_{-})$, which give no 
non-trivial geometric phase.

We emphasize that this topological property is quite different from
the familiar Aharonov-Bohm effect~\cite{aharonov}, which is topologically exact for
any finite time interval $T$. Besides, the setting of the Aharonov-Bohm effect differs from the present level crossing problem in the fact that the space is not simply connected in the case of the Aharonov-Bohm effect.

\section{Non-adiabatic phase}

We comment that the non-adiabatic phase by Aharonov and Anandan~\cite{aharonov} is based on the equivalence class (or gauge symmetry) 
which identifies all the Schr\"{o}dinger amplitudes of the form 
\begin{eqnarray}
\{e^{i\alpha(t)}\psi(t,\vec{x})\}.
\end{eqnarray}
This gauge symmetry is quite different from our hidden local
symmetry which is related to an arbitrariness of the choice of 
coordinates in the functional space.

\section{Discussion} 

The notion of Berry's phase is known to be useful in various 
physical contexts\cite{review}. Our analysis 
however shows that the  topological interpretation of 
Berry's phase associated with level crossing generally fails in 
the practical Born-Oppenheimer 
approximation where $T$ is identified with the period of the 
slower system. The notion of 
``approximate topology'' has no rigorous meaning, and it is 
important to keep this approximate topological property of 
geometric phases associated with level crossing in mind when one 
applies the notion of geometric phases to concrete physical 
processes.\\

I thank Prof. N. Nagaosa for stimulating discussions.

\end{document}